\documentclass[pop,aip,amsmath,amssymb,reprint]{revtex4-1}
\usepackage{graphicx}
\usepackage{dcolumn}
\usepackage{bm}
\usepackage{color}
\usepackage{multirow}

\renewcommand{\langle}{\left<}
\renewcommand{\rangle}{\right>}
\newcommand{\vect}[1]{{\bold #1}}
\renewcommand{\hat}[1]{{\widehat #1}}
\newcommand{\comments}[1]{}

\begin{document}


\title{Intrinsic Fluctuations of Dust Grain Charge in Multi-component Plasmas}



\author{B. Shotorban}\email{ babak.shotorban@uah.edu}
\affiliation{Department of Mechanical and Aerospace Engineering, The University of Alabama in Huntsville, Huntsville, Alabama 35899}



\begin{abstract}

A master equation is formulated to model the states of the grain charge in a general multi-component plasma, where  there are electrons and various kinds of positive or negative ions that are  singly or multiply charged. A Fokker-Planck equation is developed from the master equation through the system-size expansion method. The Fokker-Planck equation has a Gaussian solution with a mean and variance governed by two initial-value differential equations involving the rates of the attachment of ions and electrons to the dust grain.  Grain charging in a plasma containing electrons, protons and alpha particles with Maxwellian distributions is considered as an example problem. The Gaussian solution is in very good agreement with the master equation solution  numerically obtained for this problem. 

\end{abstract}


\maketitle 


\section{Introduction}
\label{Introduction}

A dust grain suspended in a plasma collects ions and electrons (plasma particles), and gains 
a net electric charge that fluctuates in time. This type of  fluctuations is 
{\em intrinsic noise} which occurs in  systems  consisting of discrete particles \cite{Kampen07}. 
The intrinsic noise is inherent  in the actual mechanism that is responsible for the evolution of 
the system, and  cannot be switched off.  The discrete particles are ions and electrons in In the grain charging system, which are attached to the grain at random times, resulting in fluctuations of its charge.  

Some phenomena observed in dusty plasmas are attributed to the intrinsic fluctuations of the grain 
charge~\cite{Fortov2004}.  One reason to so-called  dust heating, in which grains gain  
kinetic energy, is these fluctuations \cite{VKNP99,de2005stochastic,norman2011anomalous}. 
The external electric and ion forces, and the Columb forces between the dust grains are dependent
on the electric charges of the grains. As the grain charge fluctuate, so do these forces, leading to random oscillations of the grains. It is shown that if the intrinsic charge  fluctuations are strong, the grain oscillations become 
unstable \cite{MIJ99,Ivlev}.  It is also shown that the dust charge fluctuations are 
a source of dissipation and are responsible for the formation of the dust-ion-acoustic shock waves \cite{mamun2002role,DM09,alinejad2010dust}.

In the current study, the  charge fluctuations are described for a grain which is suspended in 
a plasma containing electrons and  different kinds of negative or 
positive ions that are singly or multiply charged. The assumed plasma is the most general 
case of a plasma in terms  of the  involved components.
Some applications relevant to the  problem considered here have been identified:
In the solar system, interstellar dust gains are charged by electrons, protons, and alpha particles \cite{Kimura1998,mann2010interstellar,mann2013dust}. In addition to these ions,  other highly charged ions of carbon, oxygen, and nitrogen  contribute to the charging of dust in the solar wind plasma \cite{pines2010charging,kharchenko2012charge}. 
Dust in nuclear-induced plasmas is charged by alpha particles and multiply-charged fission fragments \cite{fortov2001dust,Fortov2004}. Dust in tokamaks is charged as it collects electrons, deuterium ions, and multiply-charged impurity ions \cite{smirnov2007modelling,krasheninnikov2011dust}.

A number of approaches have been proposed  to describe the grain charge intrinsic fluctuations. They are based   on discrete stochastic modeling \cite{CG94}, master and Fokker-Planck equations \cite{MR95,MRS96,MR97,shotorban2011nonstationary,matthews2013cosmic}, or Langevin equations \cite{MR97,KNPV99,shotorban2011nonstationary,matthews2013cosmic}.  The  
 approaches given by~\citet{MR95,MRS96,MR97,KNPV99,shotorban2011nonstationary,matthews2013cosmic} are valid only for cases in which all plasma particles are singly charged. The approach given by~\citet{CG94} is valid only for cases that all contained ions have an identical charge number.  As will be seen in the following section, all these approaches are special cases of general approaches that are proposed in the present study for the description of grain charging in plasmas with various types of singly or multiply charged ions. 

The outline of the paper is as follows: In sec.~\ref{sec:master},  a master equation describing the grain charging in a general multi-component plasma through a Markovian approach is presented. In sec.~\ref{sec:FPE}, the derivation of a Fokker-Planck equation from the master equation is detailed and a solution for the derived Fokker-Planck equation is obtained. Sec.~\ref{MathForm} is continued with subsections on grain mean and variance equations,  a Langevin equation and a discrete stochastic model that can be utilized to simulate the time advancement of the fluctuating grain charge.  In sec. \ref{sec:Maxwell}, results obtained  by approaches given in sec.~\ref{MathForm} to describe intrinsic charge fluctuations of grains suspended in a Maxwellian plasma containing electrons, protons, and alpha particles are presented. In sec. \ref{sec:conclusions}, conclusions are drawn.

\section{Mathematical Formulation}
\label{MathForm}

\subsection{Master Equation}
\label{sec:master}

The differential form of the Chapman-Kolmogorov equation, known as the master 
equation~\cite{Kampen07}, is used  for the Markovian description of dust 
grain charging:

\noindent
\begin{equation}
{dP(Z,t)\over dt}=\sum_{Z^\prime} W(Z|Z^\prime)P(Z^\prime,t)-W(Z^\prime|Z)P(Z,t),
\label{eq:Master}
\end{equation}

\noindent
where $Z$ indicates the net number of  elementary charges collected on 
the grain, $P(Z,t)$ is the probability density function {{\color{black}(PDF)} of the state being at $Z$ at time $t$, 
and $W(Z'|Z)$ is the transition probability per unit time from state $Z'$ to state $Z$. 
Eq.~(\ref{eq:Master}) can be regarded as a gain-loss equation for the probabilities of 
 separate elementary charge states. Here, $W(Z|Z^\prime)$ is expressed in terms of the rate 
 of the attachment of  plasma particles (electrons or ions) to the grain:

\begin{figure}[t]
\begin{minipage}[b]{\columnwidth}
\begin{center}
\includegraphics[width=.47\columnwidth, angle=-90]{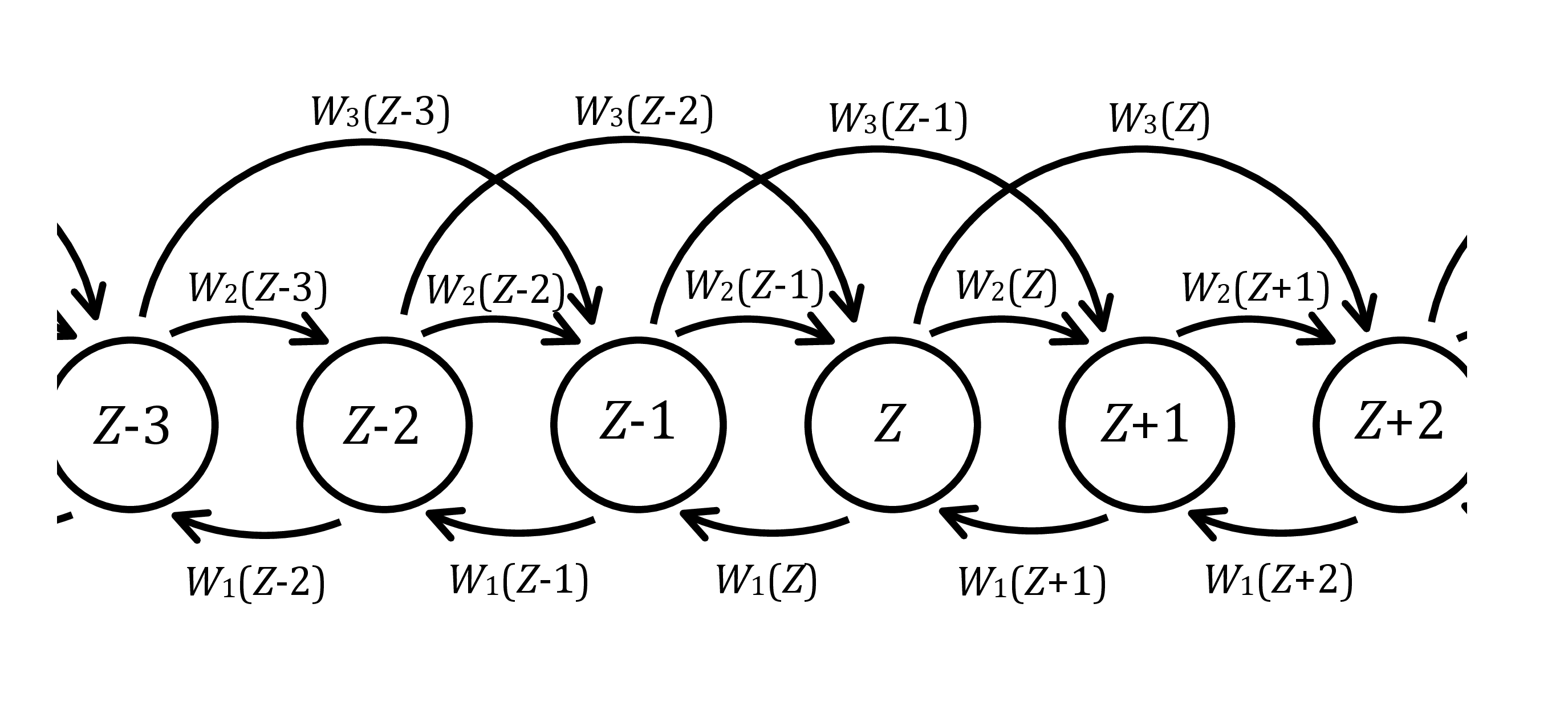}
\end{center}
\end{minipage}
\caption{States of the grain charge  in plasmas containing electrons, and singly- and 
doubly-charged positive ions, i.e., $\zeta_1=-1$, $\zeta_2=1$, and $\zeta_3=2$.  Here,
$W_1$, $W_2$ and $W_3$ indicate the rates of the attachment  of electrons, singly charged 
 ions, and doubly charged  ions, respectively.  }
\label{fig:stateDiagram}
\end{figure}

\noindent
\begin{equation}
W(Z|Z^\prime)=\sum_{j=1}^N W_j\left({Z^\prime}\right)\delta_{\zeta_j,Z-Z^\prime}, 
\label{eq:oneStepTransition}
\end{equation}

\noindent
where $\delta_{x,y}$ is the Kronecker delta,  $\zeta_j$ indicates the charge number {\color{black}(valence)} of the $j$-th plasma component particle,
 $W_j(Z')$ is the rate of the attachment of the $j$th-component plasma particles to  the grain, and $N$ is the total number of the plasma components.  The rates of attachment of plasma particles in the collisional charging of a grain are correlated with the their velocity distribution (see Appendix \ref{sec:rates}).
It is noted that the electric current of each component to the dust grain can be  calculated by 

\begin{equation}
I_j(Z)=\zeta_jW_j(Z).
\label{eq:current}
\end{equation}

\noindent

Sometimes  in addition to the direct impingement  of plasma particles
onto the grain, there are  other mechanisms such as the ultraviolet induced photoemission of  
electrons and the secondary emission of electrons \cite{RevModPhys.81.25} that contribute to 
the grain charging. If these mechanisms are also present in the charging of the grain, 
their associated rates   can be readily incorporated into eq.~(\ref{eq:oneStepTransition}).  
Then, for instance, $W_{N+1}(Z)$ and $W_{N+2}(Z)$  can represent the rate of the 
detachment of electrons  from the grain by photoemission and secondary emission, respectively.

Having substituted for $W(Z|Z^\prime)$ in  eq.~(\ref{eq:Master})  from eq.~(\ref{eq:oneStepTransition}), one obtains: 

\noindent
\begin{equation}
{dP(Z,t)\over dt}=\sum_{j=1}^NW_j(Z-\zeta_j)P(Z-\zeta_j,t)-W_j(Z)P(Z,t).
\label{eq:oneStepMaster}
\end{equation}

\noindent
Fig.~\ref{fig:stateDiagram} shows a diagram for the state of the grain charge according to 
this equation for a plasma with three components including electrons, singly charged positive ions, and doubly charged positive ions. 
{\color{black} The master equation given by \citet{MR95,MRS96,MR97,shotorban2011nonstationary} for grain charging  is a special case of eq.~(\ref{eq:oneStepMaster}), applicable for  plasmas containing only 
singly-charged components}. 
From eq.~(\ref{eq:oneStepMaster}), an equilibrium charge distribution is obtained  at the stationary state  when $t\to\infty$ and $dP/dt$ vanishes: 
\begin{equation}
\sum_{j=1}^NW_j(Z-\zeta_j)P_s(Z-\zeta_j)-P_s(Z)\sum_{j=1}^NW_j(Z)=0,
\end{equation}
\noindent 
where $P_s(Z)$ is the PDF at the stationary state.

The master equation~(\ref{eq:oneStepMaster}) may be shown in a matrix form by
\begin{equation}
\dot{\vect{P}}(t)=\vect{W}\vect{P}(t),
\label{eq:matrixMaster}
\end{equation}
where $\vect{P}(t)$ is a vector with elements $\vect{P}_m(t)\equiv P(m,t)$, and
$\vect{W}$ is the transition rate matrix with elements
\begin{equation}
\vect{W}_{mn}=\sum_{j=1}^NW_j(n)\delta_{\zeta_j,m-n}-\delta_{m,n} \sum_{j=1}^NW_j(m).
\label{eq:Wmn}
\end{equation}
The off-diagonal elements of $\vect{W}$ are the transition probability per unit time given in eq.~(\ref{eq:oneStepTransition}), i.e.,  $\vect{W}_{mn}=W(m|n)$ if $n\ne m$ while its diagonal elements $\vect{W}_{mm}=-\sum_{j=1}^NW_j(m)$ is the net escape rate from state $m$.  In the current work, eq.~(\ref{eq:matrixMaster}), which represents a system of ordinary differential equations, is solved numerically in order to find a solution for the master equation (\ref{eq:oneStepMaster}).  An  analytical solution to eq.~(\ref{eq:matrixMaster}) with $\vect{W}$ given in eq.~(\ref{eq:Wmn}) is not known. It is noted that only in rare cases it is possible to solve the master equations explicitly~\cite{Kampen07}.

\subsection{Fokker-Planck Equation}
\label{sec:FPE}
Because of the lack of an analytical solution to the master equation~(\ref{eq:oneStepMaster}), 
a Fokker-Planck equation with an analytical solution is developed  for $P(Z,t)$
from  eq. (\ref{eq:oneStepMaster}).  This development is achieved  through the system-size expansion method, which  is a systematic approximation approach for the expansion of 
master equations that describe systems with intrinsic fluctuations \cite{Kampen07}.
This approach has been recently employed for the expansion of the master equation of  
the grain charging in plasmas containing only electrons and singly charged ions~\cite{shotorban2011nonstationary,shotorban2012stochastic}. 

Let us assume that $\Omega$ represents the system size parameter, which is a reference
number of elementary charges in the grain charging case as will be seen later. Following an 
ansatz proposed by \citet{Kampen07}, a change of variable is performed: 

\noindent
\begin{equation}
Z= \Omega\phi(t)+\Omega^{1/2}\xi.
\label{eq:variableTransform}
\end{equation} 

\noindent
According to this equation, $Z$ is a combination of a deterministic part $\phi(t)$ 
scaled by $\Omega$, and a random part $\xi$ scaled by $\Omega^{1/2}$.
The probability density function can be then expressed as 

\begin{equation}
P(Z,t)=P(\Omega\phi(t)+\Omega^{1/2}\xi,t)=\Pi(\xi,t).
\label{eq:PtoPi}
\end{equation}

\noindent
Expanding the terms in the master equation in powers of $\Omega^{-1/2}$ and setting the coefficients of two highest 
powers of $\Omega$ to zero \cite{Kampen07}, results in the following equations:

\begin{equation}
{d\phi(t)\over dt}= a_1[\phi(t)],
\label{eq:Macroscopic}
\end{equation}

\begin{equation}
{\partial\Pi(\xi,t)\over \partial t}=-a_1^\prime[\phi(t)]{\partial \xi\Pi(\xi,t)\over\partial\xi}
+{1\over 2}a_2[\phi(t)]{\partial^2\Pi(\xi,t)\over\partial\xi^2}.
\label{eq:linearFP}
\end{equation} 

\noindent
where $a_k(x)=\Omega^{-1}\alpha_k({x\Omega})$ and $a_1^\prime(x)=\partial a_1(x)/\partial x$.
 Here, $\alpha_k({Z})$ is the $k$th  jump moment calculated by 
 $\alpha_k(Z)=\sum_{Z^\prime}(Z^\prime-Z)^kW(Z^\prime|Z)$, which by using eq.~(\ref{eq:oneStepTransition}), is simplified to 
 
\begin{equation}
\alpha_k(Z)=\sum_j\zeta_j^kW_j(Z).
\label{eq:jumpMoment}
\end{equation}

\noindent 

For $k=1$, one obtains
\begin{equation}
\alpha_1( Z )=\sum_j I_j(Z)
\label{eq:netCurrent}
\end{equation}

\noindent
where the right hand side is the net electric current from the plasma components to the grain according to 
eq.~(\ref{eq:current}).

\begin{figure}[t]
\vspace{2.7in}
\begin{center}
\includegraphics[width=\columnwidth]{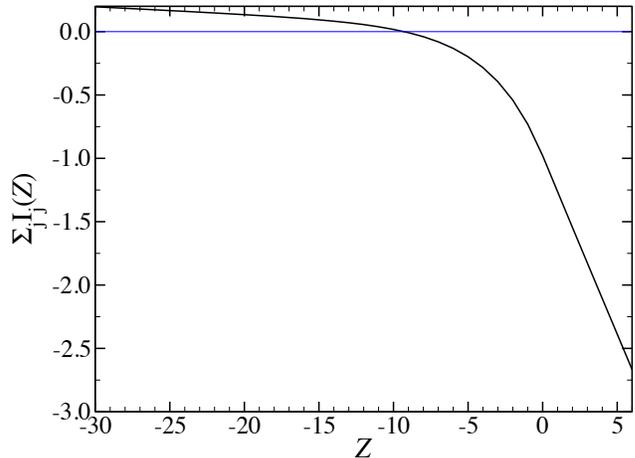}
\end{center}
\caption{Net electric current versus grain  charge for a grain with a radius of 3~nm in a multi-component plasma containing electrons $\xi_1=-1$, hydrogen ions $\xi_2=1$, and alpha particles $\xi_3=2$. Plasma components have Maxwellian distributions with 
$T_e=2\times10^4$~K, $n_e=10^4~\mathrm{m}^{-3}$, 
$T_2/T_e=T_3/T_e=1$, $n_2/n_e=0.8$, and $n_3/n_e=0.1$. The  parameters are  
for interstellar dust charging in the solar system as described by \citet{Kimura1998}.}
\label{fig:firstMoment}
\end{figure}

\subsection{Mean and Variance Equations}
Equation (\ref{eq:Macroscopic}) is referred as the macroscopic equation by \citet{Kampen07},
and  Eq.~(\ref{eq:linearFP}) is a Fokker-Planck equation with time dependent drift and diffusion coefficients. 
Since the objective is to solve for $P(Z,t)$ with an initial condition of $P(Z,0)=\delta(Z-Z_0)$, where 
$Z_0$ is the initial charge on the grain, from eq.~(\ref{eq:variableTransform}), it could be 
said  that $\phi(0)=Z_0/\Omega$, and  for all times $\langle\phi\rangle=\langle Z\rangle/\Omega$  
and $\langle \xi\rangle =0$. Thus, 
eq.~(\ref{eq:Macroscopic}), as the macroscopic equation, can be expressed by:

\begin{equation}
{d\langle Z\rangle\over d t}=\alpha_1(\langle Z \rangle),
\label{eq:meanCharge}
\end{equation}

\noindent
where  $\langle Z\rangle$ is the grain mean charge.  
According to eq.~(\ref{eq:netCurrent}), the rate of change of the mean grain charge is equal to the net current evaluated at the mean grain charge. Eq.~(\ref{eq:meanCharge}) describes the mean charge evolution with time and 
by itself is in closed form. According to this equation, the mean charge  at the stationary 
state  can be obtained by solving $\alpha_1\left(\langle Z\rangle_s\right)=0$. 
It can be shown \cite{Kampen07} that the stationary mean charge can be reached  and will be stable if 
$\alpha_1^\prime(\langle Z\rangle) < 0$ (see fig.~\ref{fig:firstMoment}).
The time evolution of the grain mean charge in a multicomponent plasma obtained by
eq.~(\ref{eq:meanCharge}) is displaced in Fig.~\ref{fig:timeEvolution}.

The Fokker-Planck equation (\ref{eq:linearFP}) has a Gaussian solution for $\Pi(\xi,t)$ 
at all $t$'s, which means $P(Z,t)$ is a Gaussian function, according to 
eq.~(\ref{eq:PtoPi}), with a time-dependent mean and variance~\cite{Kampen07}.  
The mean is governed by (\ref{eq:meanCharge}) while the variance 
is governed by the following equation:

\begin{equation}
{d\langle \tilde{Z}^2\rangle\over d t}=2\alpha_1^\prime(\langle Z \rangle)\langle \tilde{Z}^2\rangle+\alpha_2(\langle Z\rangle),
\label{eq:varianceCharge}
\end{equation}

\noindent
where $\langle \tilde{Z}^2\rangle$ is the variance of the grain charge, where  
$\tilde{Z}=Z-\langle Z\rangle$, and

\begin{equation}
\alpha_1^\prime(Z)= {\partial \alpha_1(Z) \over \partial Z} =\sum_j\zeta_j{\partial W_j(Z)\over\partial Z}.
\end{equation}

\noindent
Eq.~(\ref{eq:varianceCharge}) models the time 
evolution of the variance, and is in closed form  when coupled to Eq.~(\ref{eq:meanCharge}). 
The time evolution of $\langle Z \rangle\pm\langle \tilde{Z}^2 \rangle^{1/2}$, where
$\langle \tilde{Z}^2 \rangle$ is obtained from eq.~(\ref{eq:varianceCharge}), is displayed 
in Fig.~\ref{fig:timeEvolution}.
Also, from Eq.~(\ref{eq:varianceCharge}), the variance of the grain charge at the stationary state can be obtained as 
\begin{equation}
\langle \tilde{Z}^2\rangle_s=-\alpha_2\left(\langle Z\rangle_s\right)/2
\alpha_1^\prime(\langle Z\rangle_s).
\end{equation}
 It is noted that the Fokker-Planck equation derived 
by \citet{MR97} for the charging of a dust grain in plasmas containing only electrons and 
singly charged ions is a special case of eq.~(\ref{eq:linearFP}) where the drift and 
diffusion coefficients are given in terms of the grain charge stationary  mean 
and variance. So their Fokker-Planck is valid only for the cases that the initial charge of the dust 
grain is in the vicinity of its mean value at the stationary state.
\subsection{Langevin Equation}

Equation~(\ref{eq:linearFP}) is statistically equivalent to the following Langevin type of stochastic differential equation:

\begin{equation}
d\tilde{Z}(t)=\alpha_1^\prime\left(\langle Z\rangle\right)\tilde{Z}(t)dt+\sqrt{ \alpha_2\left(\langle Z\rangle\right)}dw(t),
\label{eq:Langevin}
\end{equation}

\noindent
where $w(t)$ is a Wiener process \cite{Gardiner04}. This equation could  have applications 
for the cases in which the dynamical behavior of the grain is of interest when charge fluctuations are 
significant. For  the case of ions being only singly charged, eq.~(\ref{eq:Langevin}) is simplified to the 
Langevin equation developed  by \citet{MR97,KNPV99} for the stationary, and by \citet{shotorban2011nonstationary} for the nonstationary charging of a grain. The time correlation of the grain charge fluctuations at the stationary state can 
be obtained as

\begin{equation}
\langle \tilde{Z}(t)\tilde{Z}(t+\tau)\rangle_s=-{\alpha_2\left(\langle Z\rangle_s\right)
\over2\alpha_1^\prime(\langle Z\rangle_s)}\exp\left[-\left |\alpha_1^\prime
(\langle Z\rangle_s)\right | \tau \right], 
\end{equation}

\noindent  based on which, the particle charging time scale can be defined as 
$\tau_\mathrm{ch}=1/|\alpha_1^\prime(\langle Z\rangle_s) |$.

\subsection{Discrete Stochastic Model}  

In the development of the Fokker-Planck equation, it is assumed that the net grain charge continuously changes over time.  
From eq.~(\ref{eq:oneStepMaster}), a model can be developed  for the discrete stochastic process of the  grain charging (the model is referred by the discrete stochastic model in this work). That is to model the time evolution of $Z(t)$ which randomly changes over  time while it only admits integer numbers. To do so, two issues must be dealt with: first, how to calculate the random time intervals at which the plasma particles are attached to  the grain; and second, how to specify the type of  the plasma particle that is attached at the time of the jump. {\color{black} That is  the moment at which the charge of the grain changes as much as the charge of the attached plasma particle.  According to the Markovian description, the jump event is based on that the time scale of the attachment process of plasma particles to the dust grain, i.e., the collision of plasma particles with the dust, is so small that the attachment is assumed to effectively instantaneously takes place.
Here, the stochastic simulation algorithm given by  \citet{gillespie2007stochastic} for simulation of  chemical kinetics with master equations similar to eq.~(\ref{eq:oneStepMaster}) is adapted for the discrete stochastic modeling of dust charging:} 

The probability, given at $Z(t)=Z$, that the attachment of a plasma particle to the grain will occur in a time between $t+\tau$ and $t+\tau+d\tau$, and that plasma particle belongs to the $j$th component of the plasma is $f(\tau,j|Z)d\tau$ where $f(\tau,j|Z)$ is a PDF obtained by

\noindent
\begin{equation}
 f(\tau,j|Z) =W_j(Z) \exp\left[-\lambda(Z)\tau\right],
\label{eq:exponential}
\end{equation}

\noindent
where $\lambda(Z)=\sum_{j=1}^NW_j(Z)$. Eq.~(\ref{eq:exponential}) implies that the time interval $\tau$ is a random number with an exponential distribution with a mean $1/\lambda(Z)$ and variance $1/\lambda(Z)^2$, and $j$ is  a random number with point probabilities $W_j(Z)/\lambda(Z)$. Thus, a single realization of $Z(t)$ can be  constructed as follows: 1.~At a given time with $Z=Z(t)$, evaluate $W_j(Z)$'s and their summation $\lambda(Z)$; 2.~Generate the time interval according to $\tau=\ln(1/r_1)/\lambda(Z)$, where $r_1$ is a random number with a uniform distribution; 3. Generate $j$, which is the smallest integer satisfying $\sum_{k=1}^jW_k(Z)>r_2\lambda(Z)$, where $r_2$ is a random number with a uniform distribution; 4.~Repeat the procedure with changing $t$ to $t+\tau$ and $Z(t)$ to $Z(t+\tau)=Z(t) + \zeta_j$.  

Fig.~\ref{fig:timeEvolution} shows a single realization of the grain charge obtained through  this discrete 
stochastic model for a plasma containing electrons, 
protons, and alpha particles. 

It is noted that the discrete stochastic model demonstrated here is simplified to the model proposed by \citet{CG94}, which was developed for cases in which there are only singly charged negative plasma particles and multiply charged positive ions with an identical charge number. 

\begin{figure}[t]
\begin{center}
\includegraphics[width=.75\columnwidth, angle=-90]{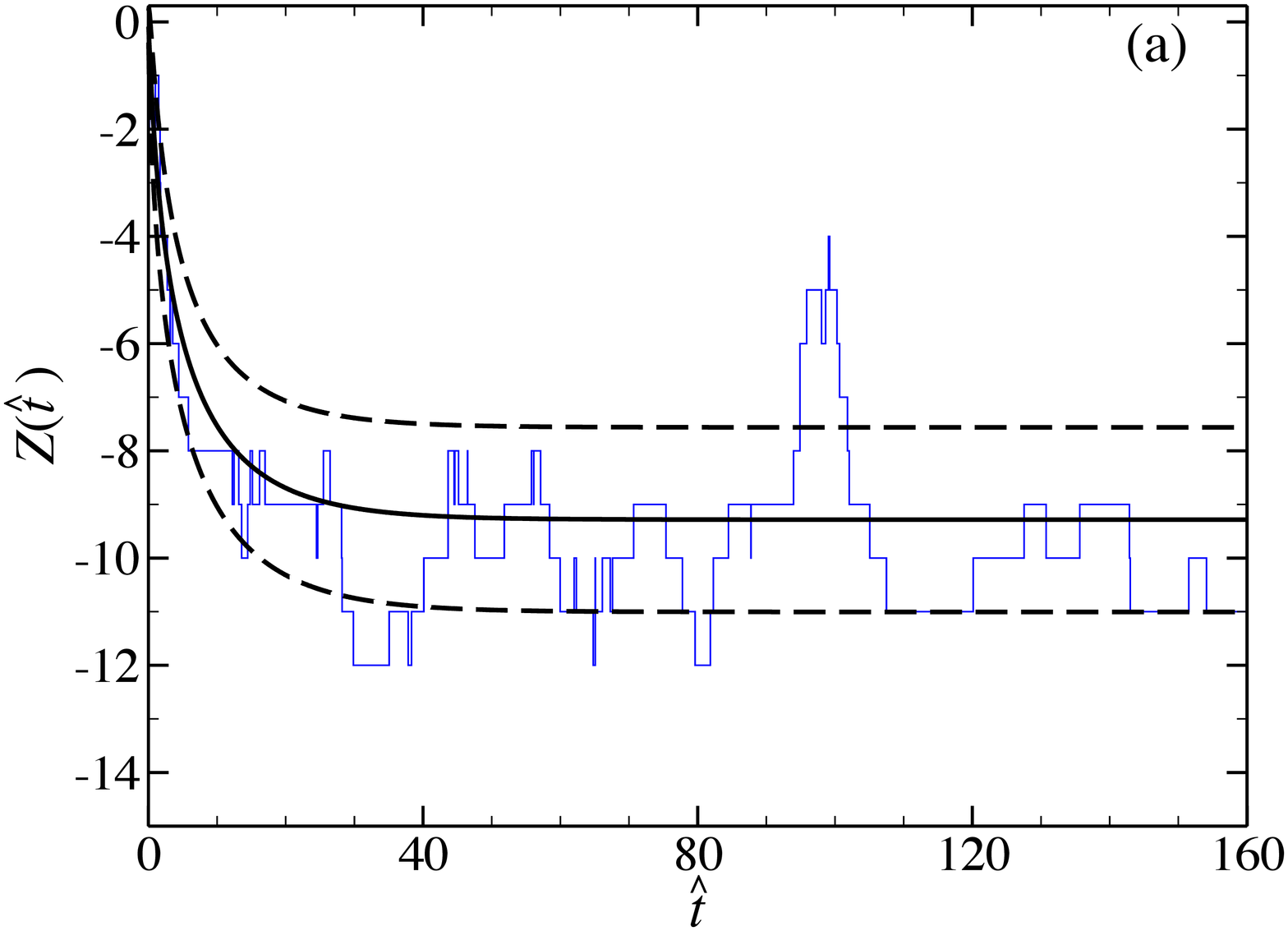}
\end{center}
\begin{center}
\includegraphics[width=.75\columnwidth, angle=-90]{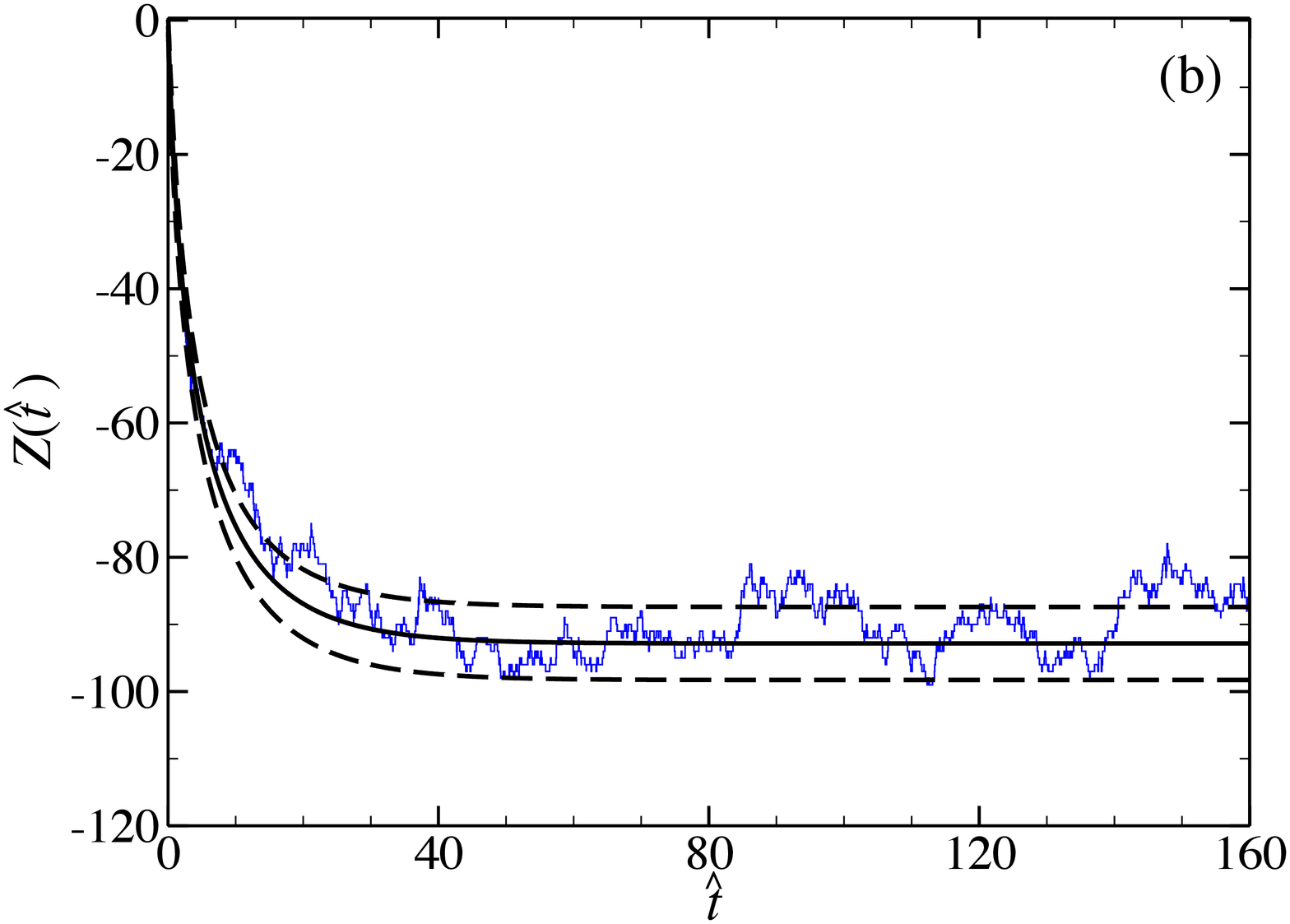}
\end{center}
\caption{Time evolution of the net number of elementary charges collected on 
a dust grain with a radius of (a) 3~nm; and (b) 30~nm, obtained by discrete stochastic model 
(step line) and the system-size-expansion method with a mean $\langle Z \rangle$ (solid line) and  
$\langle Z \rangle \pm \langle \tilde{Z}^2\rangle^{1/2}$ (dashed line). Here, $\hat{t}=\Omega^{-1}\Gamma t$ is the dimensionless time where $\Gamma$ is given by eq.~(\ref{eq:Gamma}). 
Refer to the caption of Fig.~\ref{fig:firstMoment} for the grain and  plasma properties.  }
\label{fig:timeEvolution}
\end{figure}

\begin{figure}[t]
\begin{center}
\includegraphics[width=.75\columnwidth, angle=-90]{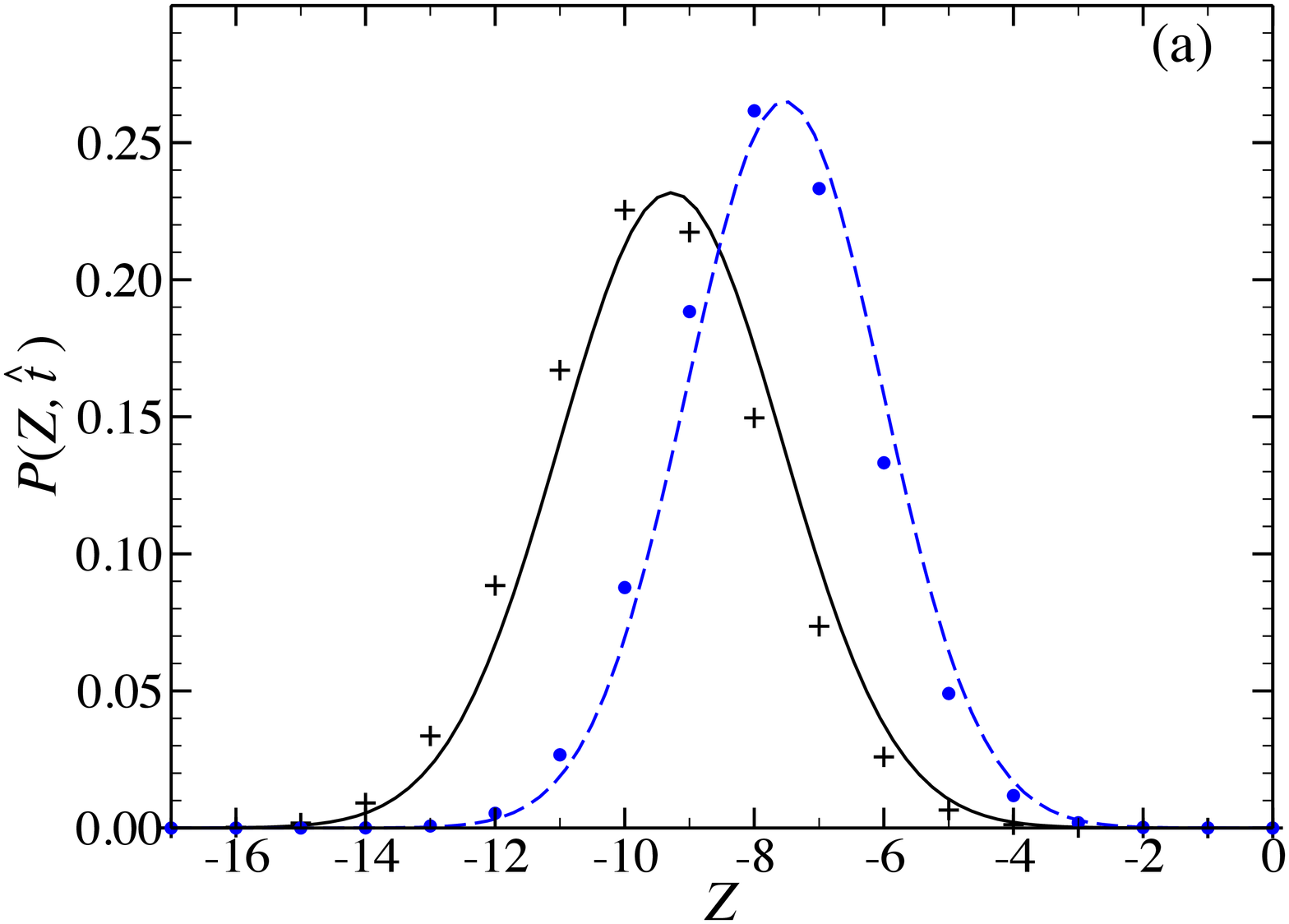}
\end{center}
\begin{center}
\includegraphics[width=.75\columnwidth, angle=-90]{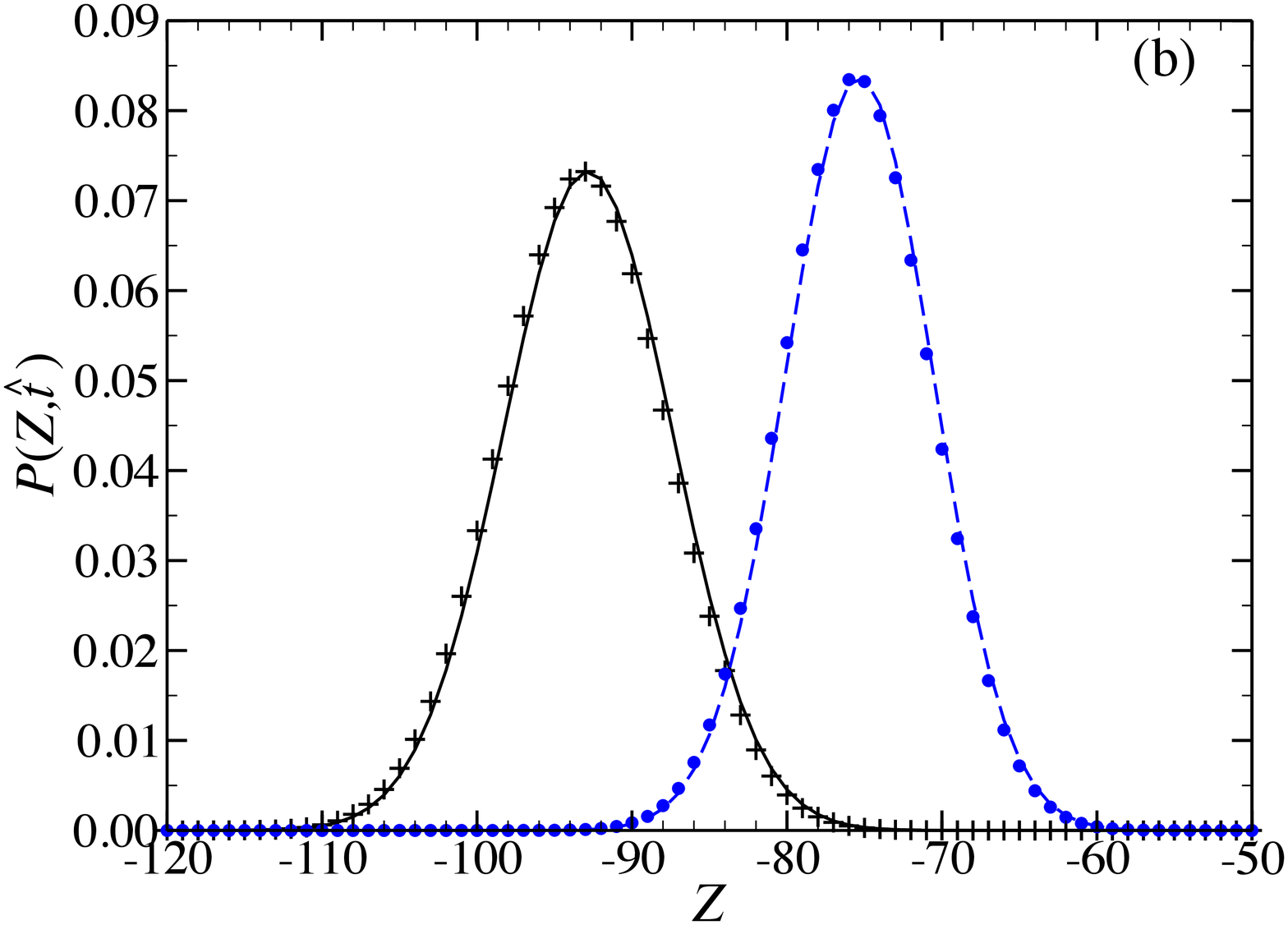}
\end{center}
\caption{Probability density function of grain charge at ${\hat t}=10$ (circle symbols and dashed lines)
and the stationary state (plus symbols and solid lines) for a grain radius of  (a) 3~nm; and (b) 30~nm,  obtained by 
solving the master equation (discrete points)  and Gaussian solution (lines) 
through the system-size expansion method. Refer to the caption of Fig.~\ref{fig:timeEvolution} 
for the grain and  plasma properties. }
\label{fig:pdf}
\end{figure}
\section{Example Problem: Grain Charging in  A Maxwellian Multi-component Plasma}
\label{sec:Maxwell}
The proposed models and approaches in this work are general and not dependent on the form of the currents. Here, as an example,  we consider  a particular case in which plasma particles follow  
Maxwellian distributions with a velocity distribution given by:

\begin{equation}
f_j(v,\theta)=n_j\left ({m_j\over 2\pi k_BT_j}\right )^{3\over 2} \exp\left(-{m_jv^2\over 2k_B T_j}\right),
\label{eq:velocityDistribution}
\end{equation}
 
\noindent 
where $n_j$ and $T_j$ are  the number density and temperature of the $j$th component particle, respectively   \cite{Kimura1998}. Substituting this equation in eq.~(\ref{eq:rate-distribution}), one obtains

\begin{equation}
W_j(Z)=  \Gamma\hat{n}_j \sqrt{\hat{T}_j\over \hat{m}_j}\times \left\{ 
\begin{array}{l l}
  1-{\zeta_jZ\over \hat{T}_j\Omega} & \quad \zeta_jZ\le 0,\\\\
\exp\left (-{\zeta_jZ\over \hat{T}_j\Omega} \right)& \quad \zeta_jZ> 0,\\ \end{array} \right.
\label{eq:rateGauss}
 \end{equation}

\noindent
where $\hat{T}_j=T_j/T_e$, $\hat{m}_j=m_j/m_e$, $\hat{n}_j=n_j/n_e$,

\begin{equation}
\Omega={{4 \pi \epsilon_0Rk_BT_e}\over e^2 },
\label{eq:omega}
\end{equation}
\begin{equation}
\Gamma=\pi R^2n_e  \sqrt{8k_BT_e\over\pi m_e } ={\Omega \omega_\mathrm{pe}R
\over\sqrt{2\pi}\lambda_{De}},
\label{eq:Gamma}
\end{equation}

\noindent
where $\lambda_{De}=\sqrt{\epsilon_0k_BT_e/n_ee^2}$ is the electron Debye length and 
$\omega_{pe}=\sqrt{n_e e^2/\epsilon_0m_e}$ is the electron plasma frequency. 
In eq.~(\ref{eq:omega}), $\Omega$ is the system size utilized in eq.~(\ref{eq:variableTransform}),
and it is a reference number of elementary charges.
According to this definition, an $R$-radius conducting sphere charged with $\Omega$ number of elementary charges will have an electric potential equal to two thirds of the mean kinetic energy of electrons of the surrounding plasma.
It is noted that in an identical plasma environment, a larger dust grain gains more charge. However, the ratio of the root mean square (rms) of charge fluctuations (charge standard deviation) to the charge mean decreases as the size of the grain increases. In fact, the system size expansion of the master equation is based on that both mean and variance of the grain charge are scaled by $\Omega$ according to eq.~(\ref{eq:variableTransform}). It is seen in 
eq. (\ref{eq:omega}) that $\Omega\propto R$ for Maxwellian distributions of plasma particles.

Figure \ref{fig:pdf} shows the probability density function of the grain charge in a plasma 
containing three components with Maxwellian distributions. Two grains with radius of 3~nm and 
30~nm are considered and the PDFs are obtained by two models. In the first model, the master 
equation (\ref{eq:oneStepMaster}) is numerically solved, using the form given in eq.~(\ref{eq:matrixMaster}), and in the second model,  the PDF is a 
Gaussian function whose mean and variance are determined by 
eqs.~(\ref{eq:meanCharge}) and (\ref{eq:varianceCharge}), respectively, via the system 
size expansion method. As can be seen, the discrepancy between the models for larger grain is 
negligible while it is more significant for smaller grain. The system size $\Omega$ for 3~nm and
30~nm grains are calculated  $3.59$ and $35.9$, respectively, so the system-size 
expansion approximation is more adequate for the larger grains. 

\section{Conclusions}
\label{sec:conclusions}
Markovian description of charging of a dust grain suspended in a general multicomponent plasma containing electrons, negative and positive singly- or multiply-charged ions was done through the formulation of a master equation. An analytical solution is lacking for the master equation. A discrete stochastic model, based on the master equation, was proposed to simulate the time evolution of the dust grain charge. Moreover, a Fokker-Planck equation was derived from the master equation through the system-size expansion method of ~\citet{Kampen07}. The Fokker-Planck equation has an analytical Gaussian solution with a mean and variance governed by two differential equations valid at both stationary and non-stationary states. As a test problem, two grains with different sizes in a plasma containing electrons, protons and alpha particles  with Maxwellian distributions were considered.  The probability density functions of the grain charge  were obtained by solving the master equation numerically and were compared against the Gaussian solutions obtained for the Fokker-Planck equation. There was very good agreement between two solutions. Results showed that the deviation from the Gaussian solution was more significant for the smaller grain. The system size, which is linearly correlated with the radius of the grain and used in the expansion of the master equation through the system-size expansion method, is smaller for the smaller grain. The approximation made in this method is less favorable for smaller system sizes. 

\acknowledgements

The author acknowledges the support through the 
2012 Junior Faculty Distinguished Research award by The University of Alabama in Huntsville.

\appendix
\section{Rates of Attachment of Plasma Particles in  Collisional Charging}
\label{sec:rates}
It is shown~\cite{dwek1992dust,Kimura1998} that the rate of the attachment of the $j$-th component plasma particle to the grain in the collisional charging of the grain is correlated to the velocity distribution of the plasma particle $f_j(v,\theta)$ by
\begin{equation}
W_j(Z)=2\pi\int_{v_0}^\infty dv\int_0^\pi d\theta \sigma_j(v,Z)f_j(v,\theta)v^3 \sin \theta,
\label{eq:rate-distribution}
\end{equation}
where $v$ is the thermal velocity of the impinging plasma particle, and $v_0$ is the minimum velocity of the impinging plasma particle, which is $v_0=\left ({\zeta_jZe^2/2\pi \epsilon_0R m_j} \right)^{1/ 2}$ if $\zeta_jZ> 0$; otherwise, $v_0=0$.
Here, $m_j$ is the mass of the $j$-th component particle and  $R$ is the radius of the grain. In equation (\ref{eq:rate-distribution}), $\sigma_j$ is the collisional cross section given by
$\sigma_j(v, Z)=\pi R^2 ( 1-{\zeta_jZe^2/ 2\pi \epsilon_0R m_j v^2}  )$.


%

\end{document}